\documentclass[11pt]{article}

\def\line#1{\hbox to \textwidth{#1}}
\catcode`\@=11

\def\thebibliography#1{\section*{REFERENCES}\list{\arabic{enumi}.}
  {\settowidth\labelwidth{#1.}\leftmargin=1.67em
   \labelsep\leftmargin \advance\labelsep-\labelwidth
   \itemsep\z@ \parsep\z@
   \usecounter{enumi}}\def\makelabel##1{\rlap{##1}\hss}%
   \def\newblock{\hskip 0.11em plus 0.33em minus -0.07em}
   \sloppy \clubpenalty=4000 \widowpenalty=4000 \sfcode`\.=1000\relax}

\def\@cite#1#2{$[{{#1\if@tempswa , #2\fi}}]$}

\newcount\@tempcntc
\def\@citex[#1]#2{\if@filesw\immediate\write\@auxout{\string\citation{#2}}\fi
  \@tempcnta\z@\@tempcntb\m@ne\def\@citea{}\@cite{%
        \@ordonner{#2}%
        \@for\@citeb:=#2\do%
    {\@ifundefined{b@\@citeb}%
        {\@citeo\@tempcntb\m@ne\@citea%
                \def\@citea{,\penalty\@m\ }{\bf ?}\@warning%
                {Citation `\@citeb' on page \thepage \space undefined}}%
        {\setbox\z@\hbox{\global\@tempcntc0\csname b@\@citeb\endcsname\relax}
     \ifnum\@tempcntc=\z@ \@citeo\@tempcntb\m@ne%
       \@citea\def\@citea{,\penalty\@m}%
       \hbox{\csname b@\@citeb\endcsname}%
     \else%
      \advance\@tempcntb\@ne%
      \ifnum\@tempcntb=\@tempcntc%
      \else\advance\@tempcntb\m@ne\@citeo%
      \@tempcnta\@tempcntc\@tempcntb\@tempcntc\fi\fi}}\@citeo}{#1}}%

\def\@citeo{\ifnum\@tempcnta>\@tempcntb\else\@citea
  \def\@citea{,\penalty\@m}%
  \ifnum\@tempcnta=\@tempcntb\the\@tempcnta\else
   {\advance\@tempcnta\@ne\ifnum\@tempcnta=\@tempcntb \else
\def\@citea{-}\fi
    \advance\@tempcnta\m@ne\the\@tempcnta\@citea\the\@tempcntb}\fi\fi}

\def\@toto{}
\newif\if@ordre 
\newcount\c@current
\newcount\c@last

\def\@ordonner#1{\global\c@last\m@ne%
                \global\@ordretrue%
                \@for\@toto:=#1\do%
                        {\@ifundefined{b@\@toto}%
                        {}%
                        {\c@current\csname b@\@toto\endcsname\relax%
                        \ifnum\the\c@current<\the\c@last\relax%
                                {\global\@ordrefalse}\fi%
                        \global\c@last\the\c@current%
                        }%
                        }%
                \if@ordre{}\else{\typeout{}%
                        \typeout{Warning: the references are not %
                         in increasing order\on@line:}%
                        \@for\@toto:=#1\do%
                        {\@ifundefined{b@\@toto}%
                        {}%
                        \typeout{\@toto:\space \@nameuse{b@\@toto}}%
                        }\typeout{}}\fi%
                }%

\catcode`\@=12

\catcode`\@=11

\newcount\c@subequation

\def\eqnarray{
\def\@eqnnum{{\reset@font\rm%
(\theequation-{\alph{subequation}})}}
\global\c@subequation=1\relax
\stepcounter{equation}\let\@currentlabel\theequation
\global\@eqnswtrue\m@th
\global\@eqcnt\z@\tabskip\@centering\let\\\@eqncr
$$\halign to\displaywidth\bgroup\@eqnsel\hskip\@centering
  $\displaystyle\tabskip\z@{##}$&\global\@eqcnt\@ne
  \hskip 2\arraycolsep \hfil${##}$\hfil
  &\global\@eqcnt\tw@ \hskip 2\arraycolsep $\displaystyle\tabskip\z@{##}$\hfil
   \tabskip\@centering&\llap{##}\tabskip\z@\cr}

\def\@@eqncr{\let\@tempa\relax
    \ifcase\@eqcnt \def\@tempa{& & &}\or \def\@tempa{& &}%
      \else \def\@tempa{&}\fi
     \@tempa \if@eqnsw\@eqnnum\global\advance\c@subequation by 1\relax
                        \fi
     \global\@eqnswtrue\global\@eqcnt\z@\cr}

\def\endeqnarray{\@@eqncr\egroup
      \global\advance\c@equation\m@ne$$\global\@ignoretrue
        \stepcounter{equation}
        \def\@eqnnum{{\reset@font\rm (\theequation)}}}

\def\Eqnarray{
\def\@eqnnum{{\reset@font\rm (\theequation)}}
\global\c@subequation=1\relax
\stepcounter{equation}\let\@currentlabel\theequation
\global\@eqnswtrue\m@th
\global\@eqcnt\z@\tabskip\@centering\let\\\@eqncr
$$\halign to\displaywidth\bgroup\@eqnsel\hskip\@centering
  $\displaystyle\tabskip\z@{##}$&\global\@eqcnt\@ne
  \hskip 2\arraycolsep \hfil${##}$\hfil
  &\global\@eqcnt\tw@ \hskip 2\arraycolsep $\displaystyle\tabskip\z@{##}$\hfil
   \tabskip\@centering&\llap{##}\tabskip\z@\cr}

\def\endEqnarray{\@@eqncr\egroup
      \global\advance\c@equation\m@ne$$\global\@ignoretrue
        \stepcounter{equation}
        \def\@eqnnum{{\reset@font\rm (\theequation)}}}

\catcode`\@=12

\font\tenmsa=msam10
\font\sevenmsa=msam7
\font\fivemsa=msam5
\font\tenmsb=msbm10
\font\sevenmsb=msbm7
\font\fivemsb=msbm5
\newfam\msafam
\newfam\msbfam
\textfont\msafam=\tenmsa  \scriptfont\msafam=\sevenmsa
\scriptscriptfont\msafam=\fivemsa
\textfont\msbfam=\tenmsb  \scriptfont\msbfam=\sevenmsb
\scriptscriptfont\msbfam=\fivemsb

\global\mathchardef\lesssim "142E

\newcommand{\slI}{\raise.15ex\hbox{$/$}\kern-.53em\hbox{$I$}}
\newcommand{\slL}{\raise.15ex\hbox{$/$}\kern-.53em\hbox{$L$}}
\newcommand{\slP}{\raise.15ex\hbox{$/$}\kern-.53em\hbox{$P$}}
\newcommand{\slR}{\raise.15ex\hbox{$/$}\kern-.53em\hbox{$R$}}
\newcommand{\slQ}{\raise.15ex\hbox{$/$}\kern-.53em\hbox{$Q$}}
\newcommand{\slK}{\raise.15ex\hbox{$/$}\kern-.53em\hbox{$K$}}
\newcommand{\slSigma}{\raise.15ex\hbox{$/$}\kern-.53em\hbox{$\Sigma$}}
\newcommand{\slcalP}{\raise.15ex\hbox{$/$}\kern-.63em\hbox{$\cal P$}}

\newcommand{\be}{\begin{equation}}
\newcommand{\ee}{\end{equation}}     
\newcommand{\bea}{\begin{eqnarray}}
\newcommand{\ena}{\end{eqnarray}}

\def\build#1\over#2{\mathrel{\mathop{\kern 0pt#1}\limits_{#2}}}

\font\tenimbf=cmmib10 at 12pt
\font\sevenimbf=cmmib10 at 7pt
\font\fiveimbf=cmmib10 at 5pt
\newfam\imbf
\textfont\imbf=\tenimbf
\scriptfont\imbf=\sevenimbf
\scriptscriptfont\imbf=\fiveimbf

\begin{document}
\begin{titlepage}
\title{\begin{center}
\end{center}
\bf{Comments on two papers by Kapusta and Wong{\footnote {and a recent unpublished paper by Wong.}}\\
 }}
\author{
P.~Aurenche$^{(1)}$, R.~Baier$^{(2)}$, T.~Becherrawy$^{(3)}$, Y.~Gabellini$^{(5)}$, F.~Gelis$^{(4)}$,\\ 
T.~Grandou$^{(5)}$, M.~Le~Bellac$^{(5)}$, B.~Pire$^{(6)}$, D.~Schiff$^{(7)}$, H.~Zaraket$^{(1)}$}
\maketitle

\begin{center}
\begin{enumerate}
\item LAPTH, 
BP110, F-74941, Annecy le Vieux Cedex, France
\item Fakult\"at f\"ur Physik, Universit\"at Bielefeld, D-33501 Bielefeld, Germany
\item Facult\'e des Sciences, Universit\'e de Nancy-1, B.P. 239,
F-54506 France
\item Brookhaven National Laboratory, Nuclear Theory, Bldg 510A, Upton, NY-11973, USA
\item INLN, 1361 Route des Lucioles, F-06560 Valbonne, France
\item CPhT, Ecole Polytechnique, F-91128 Palaiseau, France
\item LPT, Universit\'e Paris-Sud, B\^atiment 210, F-91405 Orsay, France
\end{enumerate}
\end{center}

\date{September 1, 2000}

\begin{abstract}

We critically examine recently published results on the thermal production of massive vector bosons in a quark-gluon plasma. We claim the production rate is 
a collinear safe observable.

\end{abstract}
   \vskip 4mm
\centerline{\hfill LAPTH-809/2000}
\vfill
\thispagestyle{empty}
\end{titlepage}


In two recent papers Kapusta and Wong~\cite{KapusW1,KapusW2} calculate the
order $g^2_s$ finite temperature corrections to dilepton and Z boson production
in a hot quark-gluon plasma in the limit where the dilepton or Z boson mass $M$
is large compared to the temperature $T$. This involves calculating the
two-loop QCD corrections to the imaginary part of the virtual photon or Z
self-energy diagram. The result, in both cases, is found to be
\bea
{{\rm Im}\ \Pi_{\rm{2-loop}} \over {\rm Im}\ \Pi_{\rm{1-loop}}}
&=& {10 \over 9}\ g^2_s\ {T^2 \over M^2}\ [\ \ln \Big( {2 M T \over k^2_c} \Big) + .6914\ ] \nonumber \\
&=& {10 \over 9}\ g^2_s\ {T^2 \over M^2}\ \ln \Big({4 \pi M  \over 
1.049 g^2_s T}\Big),
\label{kapwon}
\ena
where $k^2_c$ is a ``cut-off on the four momentum transfer carried by the exchanged quarks in the plasma". This cut-off is identified with ``the effective mass of a quark propagating through the plasma with a typical thermal momentum"
($m^2_{\rm eff} = g^2_s T^2 / 3$). If $m_{\rm eff}$ is taken to vanish then the two-loop expression diverges. This signals a collinear singularity
shielded by the non-vanishing thermal mass of the quark.

The result eq. (1) above is very surprising as it contradicts a series of works, published over the last twelve years, which have been ignored by the authors of refs.~\cite{KapusW1,KapusW2}. Indeed the production rate of a heavy particle in a thermal medium has been discussed several times in a variety of theories (QED, QCD, scalar theories) and has been found to be an infra-red as well as a collinear safe observable~\cite{Bella1}. These studies have been carried out in the real-time formalism using the full $2 \times 2$ matrix formalism and the imaginary part of the relevant two-point function has been calculated using the Kobes-Semenoff cutting rules~\cite{KobesS1}. All works have the common feature that intermediate steps in the calculation require the introduction of regulators which are taken to zero after all pieces are put together to construct the physical observable. The results, up to three loops in some cases, have always been found to be finite after all regulators are taken to zero.

In refs.~\cite{BaierPS2,AltheAB1,GabelGP1} the very same process as in
ref.~\cite{KapusW2} is considered namely, lepton pair production in a quark
gluon plasma (with massless quarks and gluons at zero temperature). In
\cite{BaierPS2}  two independent regulators, a gluon mass and a quark mass, are
introduced;  in \cite{AltheAB1}, instead of introducing mass regulators, the
calculation is carried out in dimensional regularisation ($n= 4 - 2
\varepsilon$); in  \cite{GabelGP1} the calculation is done in four dimensions
but a small mass is given to the gluon. In all these cases it was found that
the imaginary part of the photon two-point function, evaluated in the two-loop
approximation, is finite when the regulators vanish, indicating the absence of
infra-red and collinear singularities for this observable. In \cite{AltheB1},
the same calculation is repeated in $n$ dimensions but keeping the quark
massive: no logarithmic sensitivity to the quark mass is found in the final
result.  It is interesting to note that Cleymans and Dadic~\cite{CleymD1} have
calculated the imaginary part of the photon two-point function for a space-like
photon ($M^2 < 0$) and they also found a finite result when the regulators are
vanishing. Finally the problem of infra-red and collinear singularities was
examined in a scalar theory ($g \phi^3$ in six dimensions). Cancellation of all
singularities was found in the two-loop approximation~\cite{GrandBM1} as well
as in the three-loop approximation~\cite{GrandBP1,BellaR1}. All these results
are therefore in contradiction to eqs.~(1) (see also \cite{CleymD3,BaierPPS1} for work on related processes).

More work has been performed concerning the production of a virtual photon in a quark-gluon plasma and the ${\cal O}(g^2_s)$ terms have in fact already  been explicitely calculated ten years ago! Independently of the regularisation method used (dimensional regularisation in \cite{AltheA1} and gluon mass in \cite{GabelGP1}) the ${\cal O}(g^2_s)$ corrections have been found to be identical as it was checked in numerical calculations \cite{GabelGP1,AltheA1}.  Furthermore, in  \cite{AltheA1} a simple algebraic expression is derived in the limit of interest in refs.~\cite{KapusW1,KapusW2} ($M \gg T$)
\begin{equation}
{{\rm Im}\ \Pi_{\rm{2-loop}} \over {\rm Im}\ \Pi_{\rm{1-loop}}}
= {g^2_s \over 4 \pi^2} \Big( \ 1 \ + \ {\cal O} ({T^4 \over M^4})\
\Big),
\end{equation}
(see eq. (4.5) in \cite{AltheA1}) where the factor $1$ in the right-hand side
is just the ${\cal O}(g^2_s)$ correction factor for $e^+e^-$ annihilation into
hadrons at zero temperature\footnote{This factor 1 can be seen as a check on
the calculation and arises from the process $q \bar q G \rightarrow \gamma^*$
after summing over all allowed phase space of the initial quark, anti-quark and
gluon.} and all terms in $T^2/M^2$ vanish leaving terms in $T^4/M^4$ as leading
thermal corrections. On the other hand, a logarithmic behavior appears in the limit
$M \ll T$,
\begin{equation}
{{\rm Im}\ \Pi_{\rm{2-loop}} \over {\rm Im}\ \Pi_{\rm{1-loop}}}
= {g^2_s \over 3 \pi^2} \Big(\ (\pi^2 +8) 
\ {T^2 \over M^2}\ln{T^2 \over M^2}\ + \ {\cal O} ({T^2 \over M^2})\
\Big).
\end{equation}
It is interesting to note that the above equation has been
obtained in an explicit calculation in the bare theory~\cite{AltheA1} as well as in the limit of the effective theory~\cite{BraatP1} when the intrinsic thermal regulators (quark and gluon thermal masses) are becoming vanishingly small~\cite{AurenGKZ2}.

Not enough details are given in refs.~\cite{KapusW1,KapusW2} to locate the
origin of the error. In a recent preprint however, Wong~\cite{Wong} presents
the formalism upon which eqs.~(1) are based and he suggests that the cause of
the discrepancy between refs.~\cite{KapusW1,KapusW2} and previous works is due
to the fact that previous works (``for example \cite{AltheA1}") include a form
of resummation. This is incorrect since all previous works
\cite{BaierPS2,AltheAB1,GabelGP1,AltheB1,CleymD1,GrandBM1,GrandBP1,BellaR1,CleymD3,BaierPPS1,AltheA1}
are based on a strictly perturbative approach similarly to the approach
followed in  \cite{KapusW1,KapusW2}.

Starting from the imaginary-time formalism the author of \cite{Wong} looks for
an expression of the imaginary part of the photon two-point function. He
decomposes it into two pieces: a ``better known" $I(k)$ part and a ``not
well-known" $J(k)$ part. A physical interpretation is proposed  for each of
these pieces. The $I(k)$ part is obtained ``by putting the three internal lines
in the two-loop graphs on shell" and corresponds to physical processes
evaluated in the tree approximation. They are labeled as Compton scattering,
decay with photon emission and vector-photon fusion and they are the processes
which resemble those obtained ``by cutting rules at zero temperature". They
appear with the appropriate thermal Bose-Einstein ($f^{(+)}$ or $1+f^{(+)}$) or
Fermi-Dirac  ($f^{(-)}$ or $1-f^{(-)}$) distributions depending on whether the
particles are absorbed in the initial state or produced in the final state. The
$J(k)$ part is obtained by putting two of the three internal lines on-shell
while the uncut internal line is attributed, rather arbitrarily, the thermal
weight $1/2+f^{(+)}$ or $1/2-f^{(-)}$ for a boson or a fermion respectively.
These terms are labeled ``interference terms". 
Let us remark that the thermal weights appearing in $I(k)$ and $J(k)$
should  automatically come out from the thermal cutting rules and should not be
the result of an educated guess. Strictly adhering to the cutting rules is the
approach followed in 
refs.~\cite{BaierPS2,AltheAB1,GabelGP1,AltheB1,CleymD1,GrandBM1,GrandBP1,BellaR1,CleymD3,BaierPPS1,AltheA1}
and this leads to the correct result. There, the $I(k)$ type terms and the
$J(k)$ type terms were both included with the appropriate thermal factors and
they were labeled ``real diagrams" and ``virtual diagrams", respectively, by
analogy with the zero temperature case. 

Perhaps the best way to gauge the approach followed in \cite{Wong} is to
compare it with the retarded/advanced version of the real-time
formalism~\cite{AurenB1,EijckW1,EijckKW1,Gelis3}. Indeed this formalism allows
the construction, in the real-time approach, of Green's functions  which are
directly comparable to those obtained in \cite{KapusW1,KapusW2} from the
analytical continuation of the imaginary-time Green's functions. Expressions
for one and two-loop Green's functions are explicitely constructed
\cite{AurenB1,AurenBP1}\footnote{A brief discussion is also given on the choice
of the thermal factors of the form ($a \pm f^{(\pm)}$) where $a$ is an
arbitrary constant.}. More precisely, the very same expressions needed for the
calculation of \cite{KapusW1,KapusW2} can be found in \cite{AurenGKP2} in the
slightly more complicated case of the resummed theory (see eqs.~(24) and (29)
of \cite{AurenGKP2}).
Going back to the bare theory and comparing to the expressions in \cite{Wong}
should be an easy task.

Another subtlety in thermal calculations, which has in the past led to
discrepancies between different calculations, is related to the treatement of
the fermion effective mass in a hot medium. Thermal corrections to a fermion
propagator do not break chiral invariance~\cite{Klimo1,Kalas1,Pisar10}, {\em i.e.} the
thermally generated mass arises from a term of the form $\overline \Psi \slI
\Psi$ where $I^\mu$ is a vector-like function depending on the momentum of
fermion $\Psi$ and on the temperature. In actual calculations this leads to
distinguish~\cite{DonogH1,DonogHR1,AltheA2,BellaP1,AltheA1} between a
``kinematical mass" which incorporates the thermal effects and which appears in
the phase-space constraints and  a ``dynamical mass", the fermion mass at zero
temperature, which occurs in the evaluation of the matrix elements ({\em i.e.}
the thermal mass shift drops out in the trace evaluation). Using a scalar mass
shift $\delta m_{_{\rm Thermal}} \overline \Psi \Psi$ in the calculation would
lead to incorrect estimates of the production rate~\cite{AltheA2}.


\begin{thebibliography}{99}

\bibitem{KapusW1}
J.I. Kapusta and S.M.H. Wong, Phys. Rev. {\bf D} {\bf 62} 037301 (2000).

\bibitem{KapusW2}
J.I. Kapusta and S.M.H. Wong, Phys. Rev. {\bf C} {\bf 62} 027901 (2000).

\bibitem{Bella1}
M. Le~Bellac, Thermal Field Theory, Cambridge Monographs on Mathematical Physics, 1996; chap. 10 is entirely devoted to the problems considered here and contains an extensive list of references.

\bibitem{KobesS1}
R.L. Kobes, G.W. Semenoff, Nucl. Phys. {\bf B} {\bf 260}, 714 ({1985});
Nucl. Phys. {\bf B} {\bf 272}, 329 ({1986}).

\bibitem{BaierPS2}
R. Baier, B. Pire, D. Schiff, Phys. Rev. {\bf D} {\bf 38}, 2814 ({1988}).

\bibitem{AltheAB1}
T. Altherr, P. Aurenche, T. Becherrawy, Nucl. Phys. {\bf B} {\bf 315}, 436 ({1989}).

\bibitem{GabelGP1}
Y. Gabellini, T. Grandou, D. Poizat, Ann. of Phys. {\bf 202}, 436 ({1990}).

\bibitem{AltheB1}
T. Altherr, T. Becherrawy, Nucl. Phys. {\bf B} {\bf 330}, 174 ({1990}).

\bibitem{CleymD1}
J. Cleymans, I. Dadic, Z. Phys. {\bf C} {\bf 42}, 133 ({1989});
Z. Phys. {\bf C} {\bf 45}, 57 ({1989}).
 
\bibitem{GrandBM1}
T. Grandou, M. Le Bellac, J.L. Meunier, Z. Phys. {\bf C} {\bf 43}, 575 ({1989}).

\bibitem{GrandBP1}
T. Grandou, M. Le Bellac, D. Poizat, Phys. Lett. {\bf B} {\bf 249}, 478 ({1990}); Nucl. Phys. {\bf B} {\bf 358}, 408 ({1991}).

\bibitem{BellaR1}
M. Le Bellac, P. Reynaud, Nucl. Phys. {\bf B} {\bf 380}, 423 ({1992}).

\bibitem{CleymD3}
J. Cleymans, I. Dadic, Phys. Rev. {\bf D} {\bf 47}, 160 ({1992}).

\bibitem{BaierPPS1}
R.~Baier, E.~Pilon , B.~Pire, D.~Schiff,
Nucl. Phys. {\bf B} {\bf 336} 157 ({1990}).

\bibitem{AltheA1}
T. Altherr, P. Aurenche, Z. Phys. {\bf C} {\bf 45}, 99 ({1989}).

\bibitem{BraatP1}
E. Braaten, R.D. Pisarski, Nucl. Phys. {\bf B} {\bf 337}, 569 ({1990});
Nucl. Phys. {\bf B} {\bf 339}, 310 ({1990}).

\bibitem{AurenGKZ2}
P. Aurenche, F. Gelis, R. Kobes, H. Zaraket, Phys. Rev. {\bf D} {\bf 60}, 
076002 ({1999}).

\bibitem{Wong}
S.M.H. Wong, hep-ph/0007212.

\bibitem{AurenB1}
P. Aurenche, T. Becherrawy, Nucl. Phys. {\bf B} {\bf 379}, 259 ({1992}).

\bibitem{EijckW1}
M.A. van Eijck, Ch.G. van Weert, Phys. Lett. {\bf B} {\bf 278}, 305 ({1992}).

\bibitem{EijckKW1}
M.A. van Eijck, R. Kobes, Ch.G. van Weert, Phys. Rev. {\bf D} {\bf 50}, 4097 ({1994}).

\bibitem{Gelis3}
F. Gelis, Nucl. Phys. {\bf B} {\bf 508}, 483 ({1997}).

\bibitem{AurenBP1}
P. Aurenche, T. Becherrawy, E. Petitgirard, hep-ph/9403320 (unpublished).

\bibitem{AurenGKP2}
P. Aurenche, F. Gelis, R. Kobes, E. Petitgirard, Z. Phys. {\bf C} {\bf 75}, 315 ({1997}).

\bibitem{Klimo1}
V.V. Klimov Yad. Fiz. {\bf 33}, 1734 {(1981)} [Sov. J. Nucl. Phys. {\bf 33}, 934 {(1981)}]; Zh. Eksp. Teor. Fiz. {\bf 82}, 336 {(1982)} [Sov. Phys. JETP {\bf 55}, 1982 {(1982)}].

\bibitem{Kalas1}
O.K. Kalashnikov, Fortschr. Phys. {\bf 32}, 525 {(1984)}. 

\bibitem{Pisar10}
R.D.~Pisarski, Nucl. Phys. {\bf A} {\bf 478}, 423c ({1989}).

\bibitem{DonogH1}
J.F. Donoghue, B.R. Holstein, Phys. Rev. {\bf D} {\bf 28}, 340 {(1983)}.

\bibitem{DonogHR1}
J.F. Donoghue, B.R. Holstein, R.W. Robinett, Ann. of Phys. {\bf 164}, 233 ({1985}).

\bibitem{AltheA2}
T. Altherr, P. Aurenche, Phys. Rev. {\bf D} {\bf 40}, 4171 ({1989}).

\bibitem{BellaP1}
M. Le Bellac, D. Poizat, Z. Phys. {\bf C} {\bf 47}, 125 ({1990}).


\end{thebibliography}
\end{document}